\documentclass[traditabstract,longauth]{aa}
\usepackage{txfonts}
\usepackage{natbib}
\usepackage[dvips]{epsfig}
\usepackage{float}
\usepackage{graphicx}
\usepackage{color}

\begin{document}

\title{The VLT-FLAMES Tarantula Survey III: A very massive star in apparent isolation from the massive cluster R136\thanks{Based on 
observations at the European Southern 
Observatory Very Large Telescope in programme 182.D-0222.}}

\author{Joachim\,M.\,Bestenlehner \inst{1}
 \and Jorick\,S.\,Vink \inst{1}
 \and G.\,Gr\"afener \inst{1}
 \and F.\,Najarro \inst{2}
 \and C.\,J.\,Evans \inst{3}
 \and N.\,Bastian \inst{4,5}
 \and A.\,Z.\,Bonanos \inst{6}
 \and E.\,Bressert \inst{5,7,8}
 \and P.\,A.\,Crowther \inst{9}
 \and E.\,Doran \inst{9}
 \and K.\,Friedrich \inst{10}
 \and V.\,H\'enault-Brunet \inst{11}
 \and A.\,Herrero \inst{12,13}
 \and A.\,de Koter \inst{14,15}
 \and N.\,Langer \inst{10}
 \and D.\,J.\,Lennon \inst{16}
 \and J. Ma\'iz Apell\'aniz \inst{17}
 \and H.\,Sana \inst{14}
 \and I.\,Soszynski \inst{18}
 \and W.\,D.\,Taylor \inst{11}
 }

\institute{Armagh Observatory, College Hill,
  Armagh BT61 9DG, United Kingdom
  \and Centro de Astrobiolog\'ia (CSIC-INTA), Ctra. de Torrej\'on a Ajalvir km-4, E-28850 Torrej\'on de Ardoz, Madrid, Spain
  \and UK Astronomy Technology Centre, Royal Observatory Edinburgh, Blackford Hill, Edinburgh, EH9 3HJ, UK
  \and Excellence Cluster Universe, Boltzmannstr. 2, 85748 Garching, Germany
  \and School of Physics, University of Exeter, Stocker Road, Exeter EX4 4QL, UK
  \and Institute of Astronomy \& Astrophysics, National Observatory of Athens, I. Metaxa \& Vas. Pavlou Street, P. Penteli 15236, Greece
  \and European Southern Observatory, Karl-Schwarzschild-Strasse 2, D87548, Garching bei M\"unchen, Germany
  \and Harvard-Smithsonian CfA, 60 Garden Street, Cambridge, MA 02138, USA
  \and Dept. of Physics \& Astronomy, Hounsfield Road, University of Sheffield, S3 7RH, UK
  \and Argelander-Institut f\"ur Astronomie der Universit\"at Bonn, Auf dem H\"ugel 71, 53121 Bonn, Germany
  \and SUPA, IfA, University of Edinburgh, Royal Observatory Edinburgh, Blackford Hill, Edinburgh, EH9 3HJ, UK
  \and Departamento de Astrof\'isica, Universidad de La Laguna, E-38205 La Laguna, Tenerife, Spain
  \and European Southern Observatory, Alonso de Cordova 1307, Casilla, 19001, Santiago 19, Chile
  \and Astronomical Institute Anton Pannekoek, University of Amsterdam, Kruislaan 403, 1098 SJ, Amsterdam, The Netherlands 
  \and Astronomical Institute, Utrecht University, Princetonplein 5, 3584CC, Utrecht, The Netherlands
  \and ESA, Space Telescope Science Institute, 3700 San Martin Drive, Baltimore, MD 21218, USA
  \and Instituto de Astrof\'isica de Andaluc\'ia-CSIC, Glorieta de la Astronom\'ia s/n, E-18008 Granada, Spain
  \and Warsaw University Observatory, Aleje Ujazdowskie 4, 00-478 Warsaw, Poland
}

\date{Received 7 April 2011 / Accepted 2 May 2011 }

\abstract{VFTS\,682 is located in an active star-forming region, at a projected distance of 29\,pc 
from the young massive cluster R136 in the Tarantula Nebula of the Large Magellanic Cloud. 
It was previously reported as a candidate young stellar object, and  
more recently spectroscopically revealed as a hydrogen-rich Wolf-Rayet (WN5h) star. 
Our aim is to obtain the stellar properties, such as 
its intrinsic luminosity, and to investigate the origin of VFTS\,682. 
To this purpose, we model optical spectra from the VLT-FLAMES Tarantula Survey 
with the non-LTE stellar atmosphere code {\sc cmfgen}, as well as the spectral energy distribution
from complementary optical and infrared photometry.
We find the extinction properties to be highly peculiar ($R_{V}$ $\sim$4.7), and obtain 
a surprisingly high luminosity $\log(L/L_{\odot})=6.5\pm0.2$, corresponding to a 
present-day mass of $\sim150 M_{\odot}$.
The high effective temperature of 
$52.2\pm2.5$kK might be explained by chemically homogeneous evolution -- 
suggested to be the key process in the path towards long gamma-ray bursts.
Lightcurves of the object show variability at the 
10\% level on a timescale of years. Such changes are unprecedented
for classical Wolf-Rayet stars, and are more reminiscent of Luminous
Blue Variables.
Finally, we discuss two possibilities for the origin of VFTS\,682: (i) the star 
either formed {\it in situ}, which would have profound implications for the formation 
mechanism of massive stars, or (ii) VFTS\,682 is a {\it slow runaway} star that 
originated from the dense cluster R136, which would make it the most massive 
runaway known to date.}

\keywords{stars: Wolf-Rayet -- stars: early-type --
  stars: atmospheres -- stars: mass-loss -- stars: fundamental parameters}
\titlerunning{VFTS III: A very massive star in apparent isolation from the massive cluster R136}

\maketitle 

\section{Introduction}
\begin{figure}
\begin{center}
\resizebox{7.0cm}{!}{\includegraphics{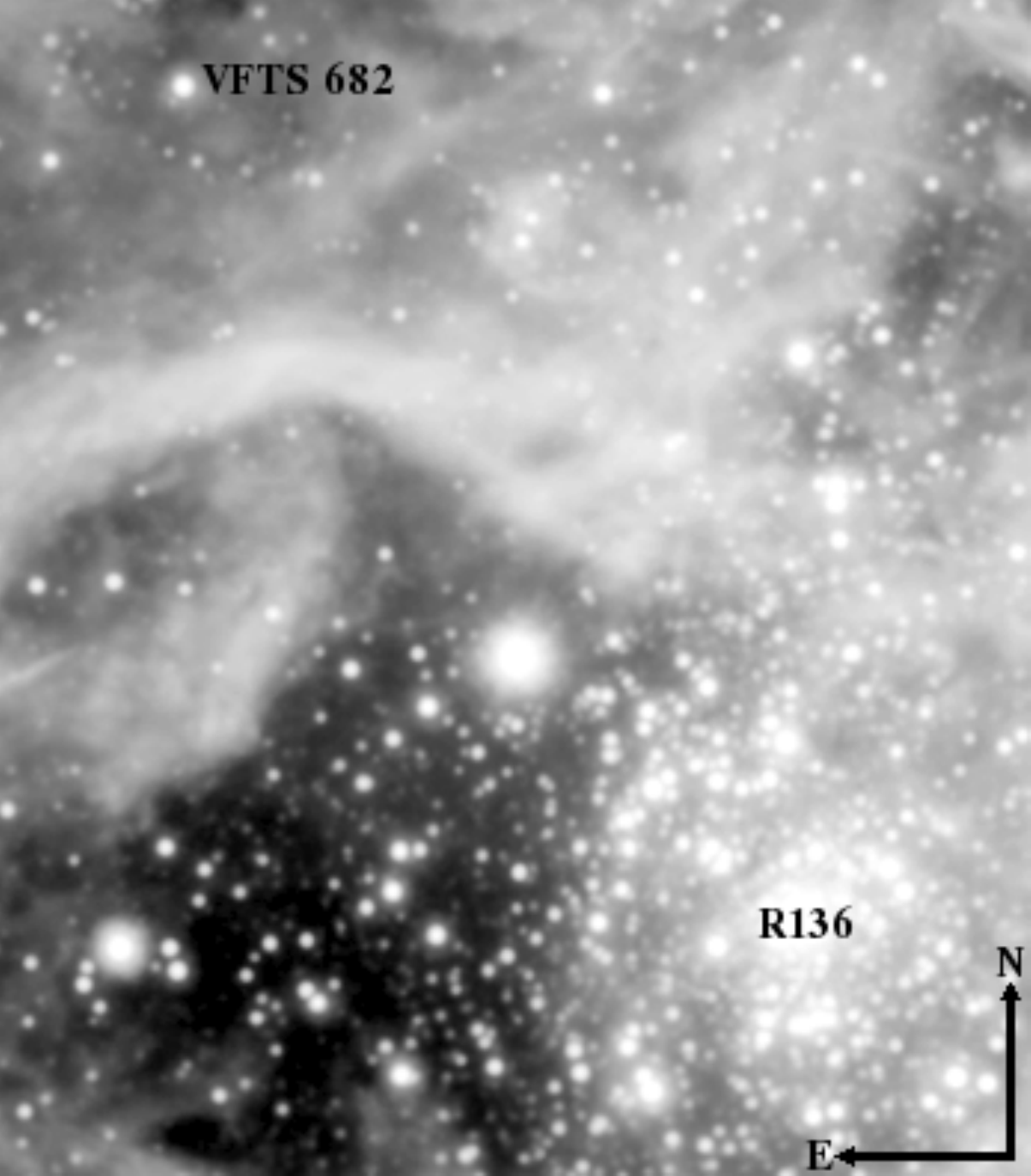}}
\end{center}
\caption{Combined $YJK_{\rm s}$ image from the VISTA Magellanic Clouds survey \citep{cioni2011}. 
The arrows correspond to 5\,pc in the Northern and Eastern directions. 
The projected distance of VFTS\,682 to the cluster R136 is 29\,pc 
(for an assumed LMC distance of 50\,kpc).}
\label{vista}
\end{figure}
Recent claims for the existence of very massive stars (VMS) of up to 300$M_{\odot}$ at the centre
of young clusters like R136 \citep{Crowther2010} seem to link the formation of such objects to
environments in the centres of massive clusters with $\sim 10^{4}-10^{5}\,M_{\odot}$. 
This finding thus appears to support massive star-formation models that invoke 
competitive accretion and possibly even merging in dense 
clusters \citep[e.g.][]{bonnell2004}. 
Such scenarios emerged after it became clear that the disk-accretion scenario -- commonly 
applied to low- and intermediate star formation -- had problems explaining the 
formation of stars with masses above 10$M_{\odot}$, as radiation pressure on dust grains would 
halt and reverse the accretion flow onto the central object \citep[e.g.][]{yorke1977}. 
However, recent multi-dimensional
hydrodynamical monolithic collapse simulations indicate that massive stars may form via disk
accretion after all \citep[e.g.][]{kuiper2010}. This illustrates that the discussion on clustered vs. isolated
massive star formation is still completely open (see \cite{deWit2005, parker2007, Lamb2010}; Bressert et al. in prep.).

In this Letter, we present evidence that VFTS\,682 (RA\,05h\,38m\,55.51s\,DEC\,-69$^\circ$\,04$'$\,26.72$''$) 
in the Large Magellanic Cloud (LMC), found in
isolation from any nearby massive, visible cluster, is one of the most massive stars known. 
The object suffers from high dust obscuration and is located in, or in the line-of-sight towards, 
an active star-forming region. 
On the basis of a mid-infrared (mid-IR) excess, \cite{gruendl2009} classified the star
as a probable young stellar object (YSO). 
The VLT-FLAMES Tarantula Survey (VFTS) \citep{evans2011} identified it as a new Wolf-Rayet (WR) star, at a projected distance
approximately 30\,pc northeast of R136. Its spectrum was classified as WN5h and is similar to 
those of the very luminous stars in the core of R136, with strong H, He\,{\sc ii}, and N\,{\sc iv} emission lines.
In this Letter, we present a photometric and spectroscopic analysis of VFTS\,682 to investigate its origin.
The star is relatively faint in the optical, which we argue is the result of significant reddening (with
Av\,$\approx$\,4.5), implying a high intrinsic luminosity and a mass of order 150\,$M_{\odot}$. The sheer 
presence of such a massive star outside R136 (and apparently isolated from any notable cluster) 
poses the question of whether it was ejected from R136 or if it was formed in isolation instead.

\section{Spectroscopic Analysis of VFTS\,682 \label{spec}}
The analysis is based on spectroscopic observations ($\lambda4000-7000$) from MEDUSA mode of VLT-FLAMES \citep{evans2011}. 
We first compared observations taken at different epochs \citep{evans2011} as to identify potential 
shifts in the radial velocity (RV) as a result of binarity. 
To this purpose, we used the N\,{\sc iv}\,$\lambda4060$ emission line. 
With detection probabilities of 96\% for a period P$<$10d,
76\% for 10d$<$P$<$1yr, and 28\% for 1yr$<$P$<$5yr, we can basically exclude that
VFTS\,682 is a short period binary. 
We found the N\,{\sc v}\,$\lambda4944$ line to be particularly useful for RV determinations, as
this line remained stable for different wind velocity laws. 
With particular emphasis on the centroid of the line we obtained
$300\pm10$\,km/s. If we give more weight to the red wing, we obtain
$315\pm15$\,km/s, with a similar number for the He\,{\sc ii}\,$\lambda4686$ line.

For the spectral analysis we use the non-LTE model atmosphere code {\sc cmfgen} 
\citep{hillier1998} which 
has been developed to provide accurate abundances, stellar 
parameters, and ionising fluxes for stars dominated by wind emission 
lines. 
We use atomic models containing H\,{\sc i}, He\,{\sc i-ii}, C\,{\sc iii-iv}, N\,{\sc iii-v}, O\,{\sc iii-vi}, Si\,{\sc iv}, P\,{\sc iv-v}, 
S\,{\sc iv-vi}, Fe\,{\sc iv-vii} and Ni\,{\sc iv-vi}. 
We made the following assumptions. 
We adopt a metallicity of half solar with respect to 
\cite{Asplund2005}, and assume N and He to have been enriched 
by the CNO process. We adopt a 12-fold enhancement of the N mass fraction. 
We employ a $\beta$ parameter of 1.6 for the wind velocity law and a wind 
volume filling factor of $f=0.25$, assuming that the 
clumping starts at 10\,km/s.
In order to estimate the error bars in the stellar temperature ($T_{\star}$) 
and mass-loss rate ($\dot{M}$), we computed a grid of models around the 
preferred values.

\begin{table}
\caption{Stellar parameters of VFTS\,682}
\label{t:paramters}
\centering
\begin{tabular}{c c} 
\hline\hline
Parameter & Value \\
\hline
$T_{\star}$					& $54.5\pm3$\,kK		\\
$T_{\rm eff}$					& $52.2\pm2.5$\,kK		\\
$\log (L/L_{\odot})$				& $6.5\pm0.2$			\\
$\log (\dot{M}/\sqrt{f}/M_{\odot}yr^{-1})$	& $-4.13\pm 0.2$		\\
$\beta$						& 1.6				\\
$f$						& 0.25				\\
$\varv_{\infty}$					&$2600\pm 200$\,km/s 	\\
Y\tablefootmark{1}				& $0.45 \pm 0.1$		\\
$M_{\rm V}$					& $-6.83 \pm 0.12$\,mag	\\
\hline
\end{tabular}
\tablefoot{\tablefoottext{1}{Y is the helium mass fraction}}
\end{table}
\begin{figure}
\begin{center}
\resizebox{\hsize}{!}{\includegraphics{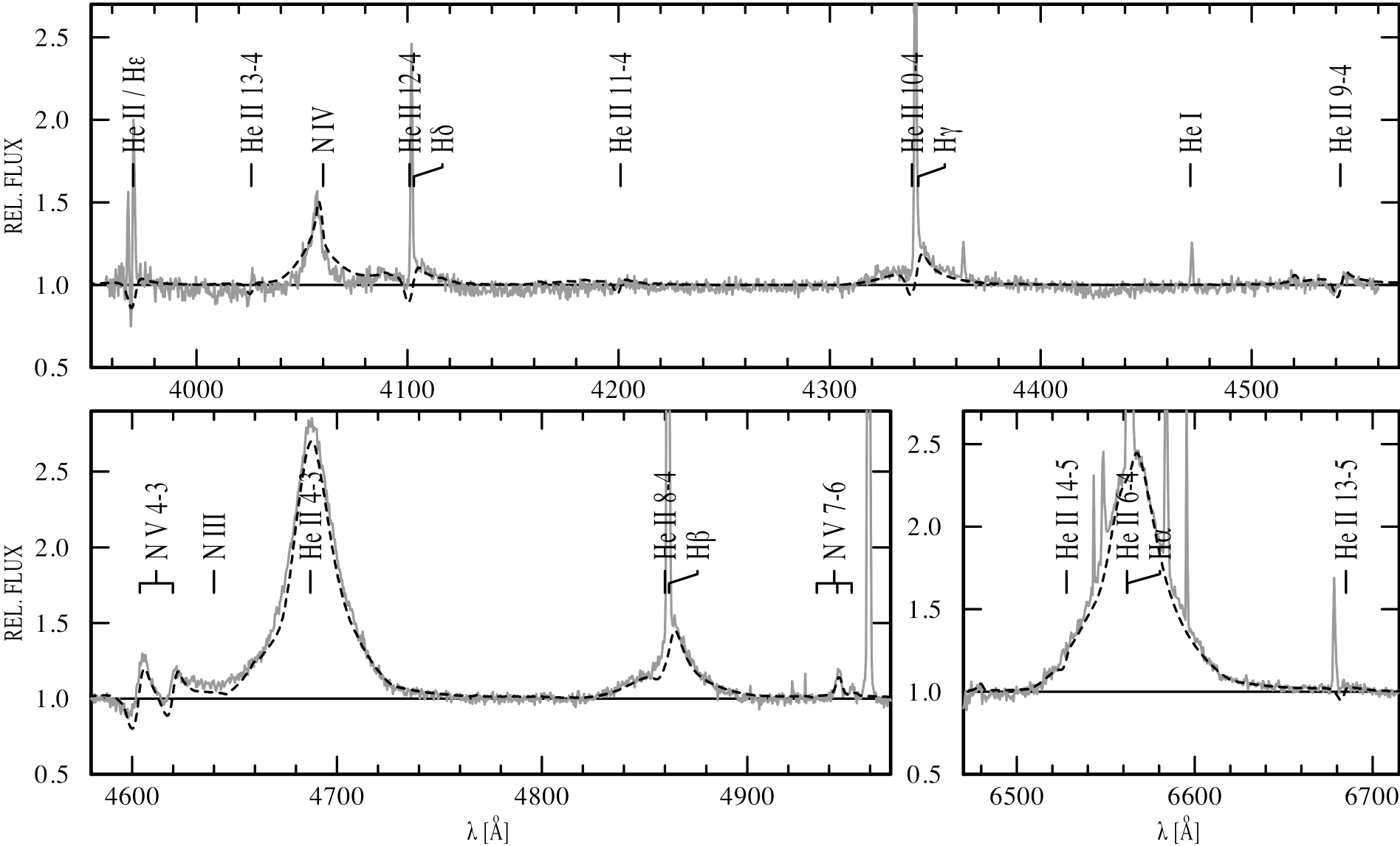}}
\end{center}
\caption{Relative flux vs. wavelength in \AA. Grey solid line: MEDUSA spectrum of VFTS\,682. Black dashed line: model spectrum.}
\label{medusa}
\end{figure}
The core temperature ($T_{\star}$) at high optical depth of $54500\pm 3000$\,K is based on the N\,{\sc iv}\,$\lambda4060$ and N\,{\sc v}\,doublet\,$\lambda4604$ and ${4620}$ 
in emission (Fig.~\ref{medusa}). N\,{\sc iv} is also sensitive to the mass-loss rate, which 
is obtained from other lines.
The hydrogen and the helium lines are used to determine both 
the helium abundance (45\% by mass) and the mass-loss rate, 
with $\log({\dot{M}/M_{\odot}yr^{-1}}) = -4.4 \pm 0.2$, where the volume filling factor corrected 
mass-loss rate is given in Table~\ref{t:paramters}. 
This value is larger than the $\log({\dot{M}}/M_{\odot}yr^{-1}) = -4.69$ obtained from the mass-loss recipe 
of \cite{vink2001} for non-clumped winds. The values can be 
brought into agreement for a volume filling factor of $f=0.1$.
In the absence of UV spectroscopy of the $\lambda1548$,~$\lambda1551$ C\,{\sc iv} doublet, 
we estimate the terminal velocity ( ${\varv_{\infty}}=2600\, \mathrm{km/s}$) 
from the broadening of the He\,{\sc ii} and $\mathrm{H_{\alpha}}$ lines. 
An overview of the stellar parameters and abundances is provided 
in Table~\ref{t:paramters}. 

The stellar parameters of VFTS\,682 are 
similar to those of R136a3 (with a $\log (L/L_{\odot})=6.58$) \citep{Crowther2010}. 
We note that there are other luminous WNh objects in 30 Dor, for instance, VFTS\,617 (=Br 88) which has the same spectral type as VFTS\,682 but with weaker emission lines. The spectral analysis of the additional 
VFTS WNh stars is currently in progress (Bestenlehner et al. in prep.).

\section{Luminosity and Reddening\label{reddening}}

\begin{figure}
\begin{center}
\resizebox{\hsize}{!}{\includegraphics{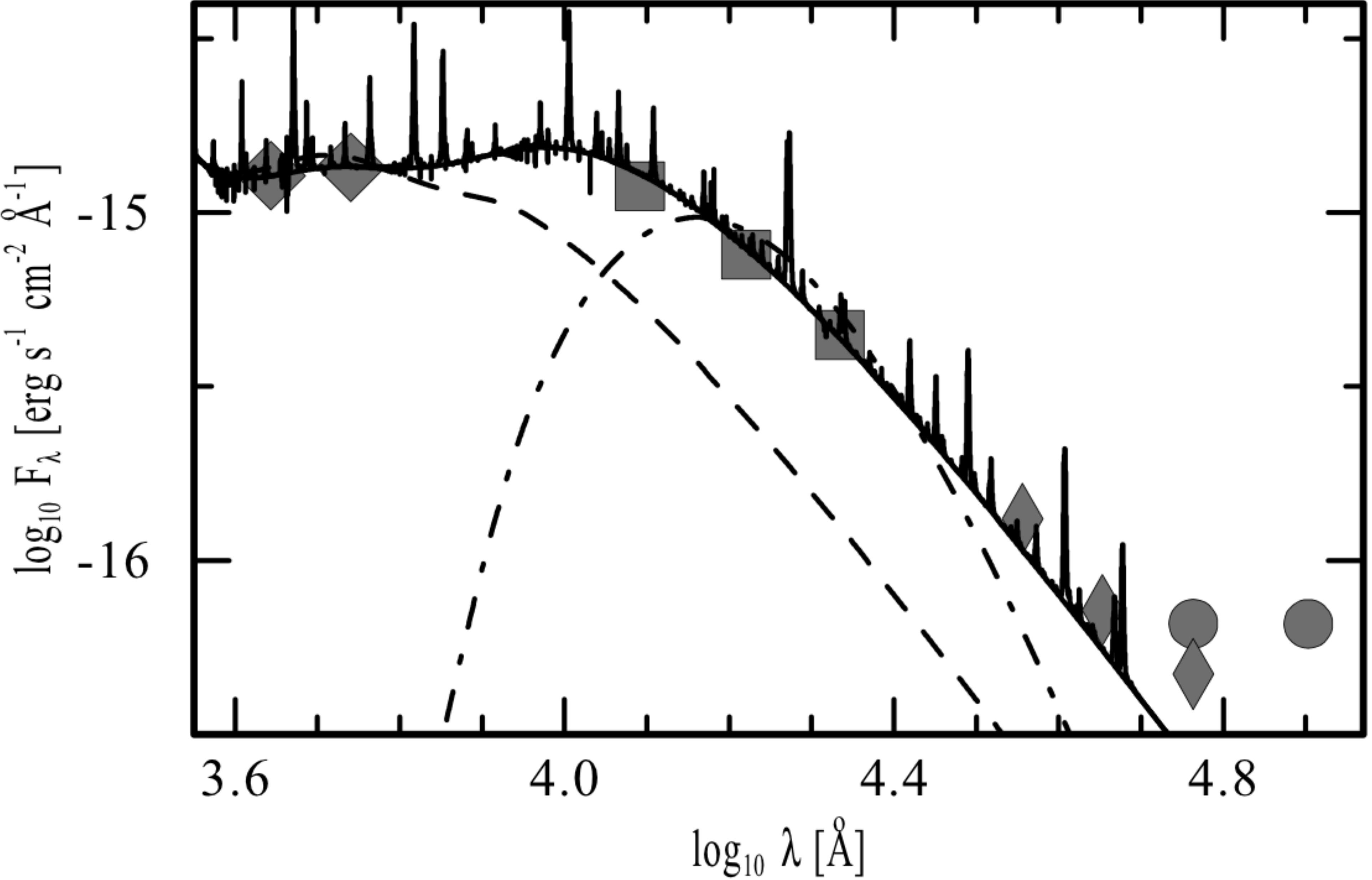}}
\end{center}
\caption{Log flux vs. log wavelength. Squares: $B,\,V,\,J,\,H,\,K_s$ photometry, Diamonds: {\sc spitzer} SAGE. Circles: 
{\sc spitzer} $5.8\,\mu m$ and $8.0\,\mu m$ from \cite{gruendl2009}. Solid line: SED with anomalous $R_{\rm V}$ parameter. Dashed line: Standard $R_{\rm V}$ parameter. Dashed-Dotted line: attempt 
to reproduce the NIR excess with a 1500 K black body. \label{sed}}
\end{figure}

In its spectral appearance, VFTS\,682 is almost 
identical to that of R136a3 in the core of the dense cluster R136. 
The luminous objects in this cluster are thought to be the most
massive stars known, with initial masses up to $300\,M_\odot$
\citep{Crowther2010}. As VFTS\,682 is apparently isolated, and shows no
signs of binarity, it offers the opportunity to study such a luminous WN5h
object in isolation.
The precise luminosity of VFTS\,682 is thus of key relevance for this work. 
Its derivation is however complicated by the large extinction.

\begin{figure}
\begin{center}
\resizebox{7.5cm}{!}{\includegraphics{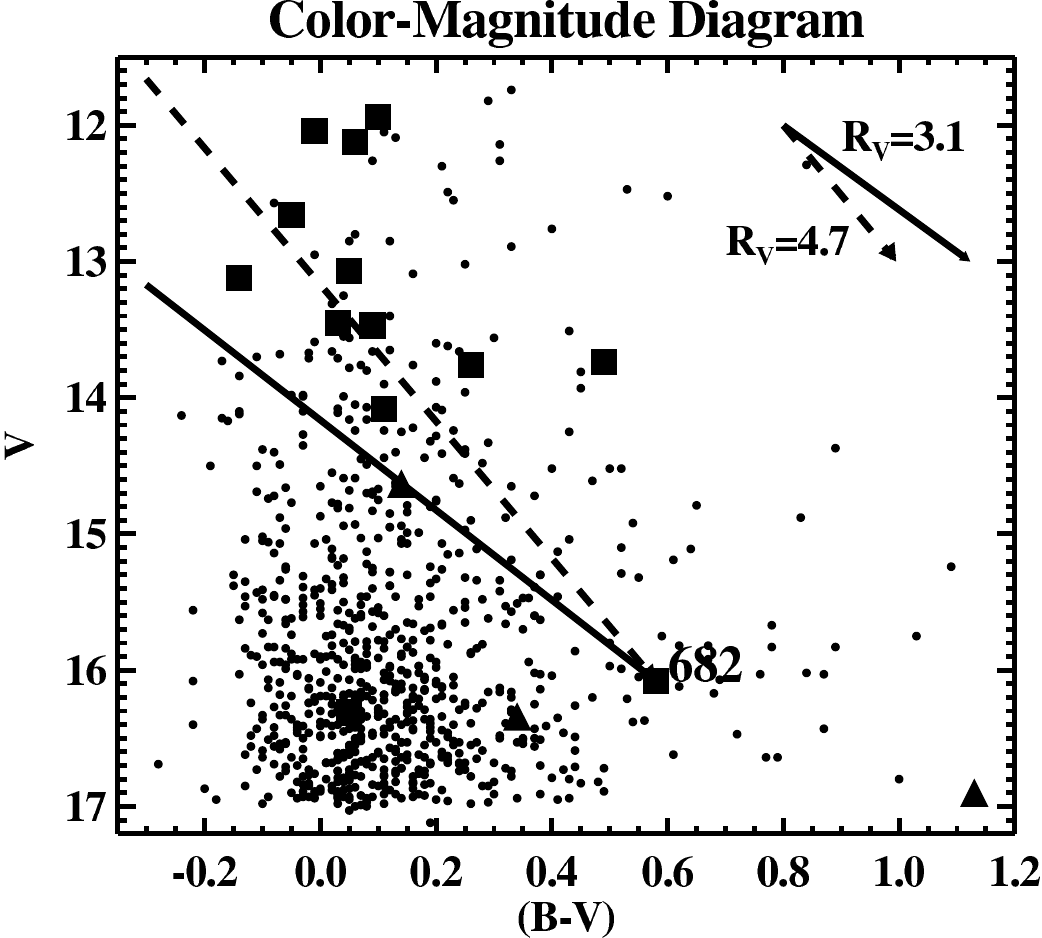}}
\end{center}
\caption{VFTS objects. 
Rectangles are WN-stars. Triangles are WC-stars. 
The arrows in the top right corner are for 
two different $R_{\rm V}$ values, with a length corresponding 
to $A_{\rm V}=1$. 
The longer lines on the left indicate the transformations 
between reddened and unreddened CMD positions.}
\label{cmd}
\end{figure}

Matching the optical photometry with a standard extinction law, i.e.,
a 'Galactic average' extinction parameter $R_{\rm V}= 3.1$, we obtain a
luminosity of $\log(L/L_\odot)=5.7\pm0.2$. With this relatively low
luminosity we can explain the optical flux distribution, but we fail
to match the observed spectral energy distribution (SED) at longer wavelengths. In the near--mid infrared (IR) 
the observed flux is $\sim 3$ times higher than the corresponding
model flux. Below, we show that we can successfully explain
this 'excess' IR emission, including its exact shape, with a
reddening parameter $R_{\rm V}=4.7$, which leads to a much higher
stellar luminosity. This effect is visualised in the colour-magnitude
diagram (CMD) of Fig.\,\ref{cmd}. We note that high $R_{\rm V}$ parameters,
that deviate from the Galactic average, are not extra-ordinary for massive stars in 
the LMC \citep[e.g.][]{bonanos2011}

\subsection{Modelling the near IR flux}

Because the extinction in the IR is very low, we can avoid the
problems resulting from the uncertain $R_{V}$ parameter, by choosing
the near-IR photometry as our flux standard.  In more detail, we
derive the luminosity of VFTS\,682 by matching $B$, and $V$ from
\citet{parker1993}, and $K_{\rm s}$ from the ``InfraRed Survey Facility (IRSF) Magellanic Clouds
Point Source Catalog'' \citep{kato2007} (see Tab.~\ref{t:phot}). For
this purpose we extract the intrinsic $B$, $V$, and $K_{\rm s}$
magnitudes from our model, using appropriate filter functions. Based
on the resulting values of $E(B-V)=0.94\pm0.02$, and $E(V-K_{\rm s})=3.9$,
we derive $R_V=A_V/E(B-V)$ for two oft-used extinction laws.

To determine $R_V$, we use the relations $R_{\rm V} = 1.1994 \times
E(V-K_{\rm s})/E(B-V) - 0.183$, as inferred from the extinction law by
\citet{cardelli1989}, and $R_{\rm V} = 1.12 \times E(V-K_{\rm s})/E(B-V) + 0.016$, 
from \citet{fit1:99}. This way, we obtain
$R_{\rm V}=4.7\pm0.1$, $A_{\rm V} = 4.45\pm0.12$, and $A_{K_{\rm s}}
= 0.55\pm0.15$, where the uncertainties refer to the differences in
the adopted extinction laws. As $A_V$ is much larger than $A_{K_{\rm s}}$, 
the derived luminosity mostly relies on $K_{\rm s}$, while
the observed values of $B$, and $V$ mostly determine $R_V$. Small
uncertainties in $B-V$ will thus mainly affect $R_V$, in a way that
the absolute values of $B$, and $V$ stay the same. As
$R_V=A_V/E(B-V)$, this introduces an anti-correlation between the
derived $R_V$, and $E(B-V)$, where $A_V$ is preserved. The resulting
luminosity is $\log(L/L_\odot)=6.5 \pm 0.2$, where the error includes
uncertainties in the stellar temperature, mass-loss rate, and
extinction. The resulting SED fit is presented in Fig.~\ref{sed}.

In an independent analysis of the available UBVIJHK photometry, using
the CHORIZOS code \citep{ma-ap2007} for CHi-square cOde for parameteRized modelIng and
characteriZation of phOtometry and Spectrophotometry, we obtain
$E(B-V)=0.98\pm0.03$, and $R_V=4.55\pm0.17$, which, due to the
aforementioned anti-correlation between $E(B-V)$ and $R_V$, results in
an almost unchanged value of $A_V=4.46\pm0.06$. The resulting
luminosity, and high $R_V$ are thus confirmed.

Table\,~\ref{t:phot} gives a summary of the available photometry for VFTS\,682. 
In analogy to our previous findings, the differences between the optical photometry of \cite{evans2011}, and 
\cite{parker1993} mainly affect $R_{\rm V}$, but not the derived luminosity.
Based on the WFI photometry by \cite{evans2011}, we obtain $R_{\rm V} = 5.2 \pm 0.1$ with $E(B-V) = 0.84\pm0.02$. 
The near-IR photometry however has a direct influence on our results.
Based on the fainter Two Micron All Sky Survey (2MASS) photometry the derived luminosity decreases by about 
0.1dex to $\log(L/L_\odot)=6.4$. 
In this work, we use the IRSF values, as they connect better to the 
observed {\sc spitzer} MIR photometry.

If we alternatively adopt the standard reddening parameter $R_{\rm V}=3.1$, and attribute the 
NIR excess to a second component, such as a cool star 
and/or a warm (1000-2000\,K) dust component (see Fig.~\ref{sed}), we obtain 
a luminosity of only $\log(L)=5.7 \pm 0.2$. 
We note that just one extra component (cool star and/or dust black body)
cannot fit the entire SED slope, which implies that we would need 
a multitude of additional components. These would need to 
conspire in such a way as to precisely follow our SED. 
Although we cannot disprove such a configuration, it would be 
highly contrived.

To summarise, the most likely luminosity of VFTS\,682 is $\log(L)=6.5
\pm0.2$, which would support the high luminosities that have been
derived by \cite{Crowther2010} for the VMS in the core of R\,136.

\subsection{The MIR excess \label{mir}}

We also investigate the issue of whether VFTS\,682 has a mid-IR excess 
in {\sc spitzer} data. 
We compared two sources of MIR photometry, the ``Surveying the Agents of a Galaxy's Evolution (SAGE) IRAC Catalog'' for point 
sources \citep{meixner2006} as well as the recent YSO catalog of \cite{gruendl2009}. 
In contrast to the SAGE IRAC Catalog, \cite{gruendl2009} included spatially more extended sources. 
The MIR photometry data from both sources are listed 
in Table~\ref{t:phot}. \cite{gruendl2009} detected a mid-IR excess 
at 5.8 and 8.0\,$\mu m$ and categorised VFTS\,682 as a YSO candidate. 
We are not able to confirm the MIR excess on the basis of the SAGE IRAC
Catalog photometry, but the difference in the two data sources may be 
explained in the following way.

The MIR point-source data of \cite{meixner2006} basically follow the 
slope of our optical-to-NIR fit continued into the MIR, whilst the larger 
scale MIR data of \cite{gruendl2009} show a clear MIR excess.
Whereas the MIR excess could be consistent with an extended circumstellar
shell, such as recently detected in {\sc spitzer} 24\,micron images
of the Galactic Centre WNh star WR102ka \citep{barniske2008}, there is no 
evidence for such a shell (or bowshock) in {\sc spitzer} 8 micron 
images of VFTS\,682 at similar angular scales as for WR102ka. 
However, given the resolution limit of $\sim 2$\,arcsec the IR excess 
might be due to an unresolved circumstellar shell with a diameter of $<0.5$\,pc.

\begin{table}
\caption{Apparent Magnitude of VFTS\,682}
\label{t:phot}
\centering
\begin{tabular}{c c c c c} 
\hline\hline
				& $U$		&	$B$	& $V$		&	$I$	\\
\hline
\cite{evans2011}		& --		&16.66		& 16.08		& 		\\
\cite{parker1993}		& 16.40		&16.75		& 16.11		&		\\
\cite{denis2005}		&		&		&		&	14.89	\\
\hline\hline
				& $J$		& $H$		& $K_s$	&		\\
\hline
IRSF\tablefootmark{1}		& 13.55		& 12.94		& 12.46		&		\\
2MASS				& 13.73		& 13.08		& 12.66		&		\\

\hline\hline
	 			& $3.6\mu m$	& $4.5\mu m$	& $5.8\mu m$	& $8.0\mu m$	\\
\hline
SAGE				& 11.73		& 11.42		& 10.84		& --		\\
\cite{gruendl2009}		&11.69		& 11.37		&10.48		&9.15		\\
\hline
\end{tabular}
\tablefoot{\tablefoottext{1}{Transformed to the 2MASS system.}}
\end{table}

\section{Variability on timescales of years \label{var_title}}

\begin{figure}
\begin{center}
\resizebox{\hsize}{!}{\includegraphics[angle=-90]{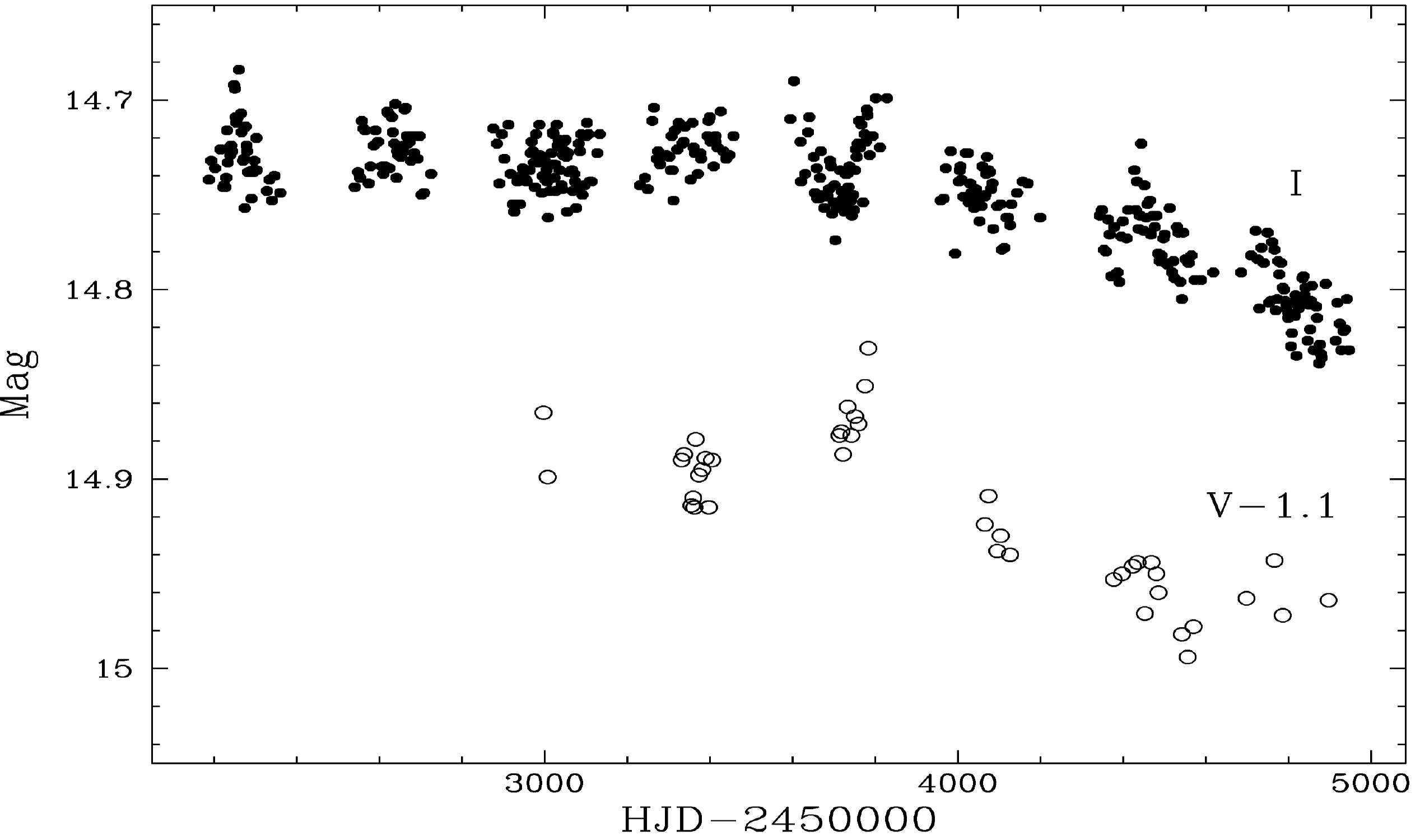}}
\end{center}
\caption{OGLE III V and I lightcurves for VFTS 682 during 2001-2009.} 
\label{var}
\end{figure}

To study the stability of VFTS\,682 we show Optical Gravitational
Lensing Experiment (OGLE-III) lightcurves \citep{udalski2008} in V and I in Fig.\,\ref{var}. 
On top of the short-term jitter, the object 
clearly dimmed by $\sim$0.1 mag in both the V and I band during the 
years 2006-2009.  
Furthermore, the object shows an increase in the K-band 
by $\sim$0.15\,mag during 2010 (Evans et al. in prep.). 
This kind of long-term variability is unprecedented for Wolf-Rayet stars, 
and is more characteristic for Luminous Blue Variables (LBVs). However, we note that 
the nature of these changes is not the same as those of  
bona-fide LBVs with S Doradus cycles of 1-2\,mag variations 
\citep{humphreys1994}.

\section{Discussion on the origin of VFTS\,682 \label{prop}}

VFTS\,682 is located in an active star-formation region of 30 Dor \citep{johansson1998}. 
It is not anywhere close to the core of R136, or -- as far as we can tell -- to any other nearby cluster. This might be
considered a surprise as VMS are normally found in the cores of large clusters, such as the Arches cluster or R136. 
By contrast, VFTS\,682 is rather isolated at a 
projected distance of 29\,pc from R136. This raises the question of whether the object
formed {\it in situ} or is a {\it slow} runaway object from R136 instead.  

In order to address these issues, we provide some velocity estimates. 
The measured RV shift of VFTS\,682 is $v_r\approx300$\,km/s, which is somewaht higher than 
the average velocity in 30\,Dor ($\sim270\pm$10\,km/s, Sana et al. in prep, \citeauthor{bosch2009}~\citeyear{bosch2009}).
If VFTS\,682 is a runaway from R136, it would require a tangential 
velocity of 30\,km/s to appear at a projected distance of 30\,pc within 
approximately 1\,Myr. Together with an RV offset of $\sim$30\,km/s, VFTS\,682 
would then require a true velocity of $\sim$40 km/s, i.e. at the lower 
range of velocities for classical OB runaway stars \citep[e.g.][]{philp1996}. 
Still, this would make it the most massive runaway star known to 
date.

In case the object is a runaway star, a bow shock might potentially be visible 
as VFTS\,682 is surrounded by dust clouds.
Whilst there is currently no evidence for a bow shock, one of the 
possible explanations for the MIR excess 
is the presence of a bow-shock \citep[e.g.][]{gvaramadze2010}. 
Alternative explanations for the MIR excess may involve its (line-of-sight) 
association with an active star forming region, or a nebula formed by 
vigorous mass loss since the object started to burn hydrogen in its core.
In this context, it may or may not be relevant that VFTS\,682 shows slow 
photometric variations suggestive of LBVs that may suffer from mass ejections.   

The most probable luminosity of VFTS\,682 is $\log(L)=6.5 \pm0.2$.
The question is what stellar mass the object corresponds to, and what is its
most likely evolutionary age. 
The high temperature places the object right on the zero-age main sequence 
(ZAMS), which can be best understood as a result of chemically 
homogeneous evolution (CHE). Such an evolution has also been proposed for 
two WR stars in the SMC \citep{martins2009}. 
Note that this type of evolution has been suggested 
for the production of long gamma-ray bursts \citep{yoon2005}. 
In order to obtain a meaningful mass limit, we employ the recent 
mass-luminosity relationships by \cite{graefener2011} 
for homogeneous hydrogen burners. 
Utilising the derived He abundance (Y=0.45), the most likely 
present-day mass is $\sim$150$M_{\odot}$. 
In detailed stellar evolution models of \cite{brott2011} and Friedrich 
et al. (in prep), CHE is achieved through rapid rotation, with
$v_{\rm rot}^{\rm init} > 200$km/s. 
A value of Y=0.45 is obtained at an age of 1-1.4 Myrs
and the initial mass would be of $\sim 120-210M_{\odot}$, 
where the mass range is defined through the uncertainty in the luminosity.

Finally, from our analysis it is clear that VFTS\,682 is a very massive object. 
It is often assumed that such massive objects can only be formed in dense cluster
environments, where they are normally found.
The apparent isolation of VFTS\,682 may thus represent an interesting 
challenge for dynamical ejection scenarios and/or massive star formation theory.
\vspace{0.1cm}

\bibliographystyle{aa}
\bibliography{17043ref}
\end{document}